\documentclass{article}
\usepackage{spconf}

\makeatletter
\renewcommand{\paragraph}[1]{%
  \@startsection{paragraph}{4}{\z@}%
    {-0.5\baselineskip}{-1em}%
    {\normalfont\normalsize\bfseries}*{#1}%
}
\makeatother

\usepackage{amsmath}
\usepackage{bm}
\usepackage{graphicx}
\usepackage{times}
\usepackage{epsfig}
\usepackage{amssymb}
\usepackage{booktabs}
\usepackage{fancyhdr}
\usepackage{multicol}
\usepackage{nicefrac}
\usepackage{transparent}
\usepackage{watermark}
\usepackage{pifont}

\usepackage{booktabs}
\usepackage{multirow}
\usepackage{array}
\usepackage{siunitx}
\usepackage{algorithm, algorithmic}
\usepackage{makecell}
\usepackage{tabularx}   %
\usepackage{siunitx}

\usepackage[dvipsnames]{xcolor}
\definecolor{natblue}{HTML}{0060BF}
\definecolor{natorange}{HTML}{BF6000}
\definecolor{natred}{HTML}{BF0000}
\definecolor{natpurple}{HTML}{78308F}
\definecolor{natgreen}{HTML}{308F30}
\definecolor{natteal}{HTML}{308F8F}

\usepackage[most]{tcolorbox}
\newcommand{\mathcbox}[2][yellow]{%
    \tcbhighmath[
        enhanced, colback=#1, frame hidden, boxrule=0pt, arc=1pt,
        boxsep=0pt, left=1pt, right=1pt, top=2pt, bottom=2pt
    ]{#2}%
}

\usepackage{enumitem}
\setlist[itemize]{noitemsep, topsep=-\parskip, parsep=0pt}
\setlist[enumerate]{noitemsep, topsep=-\parskip, parsep=0pt}

\DeclareMathOperator{\ReLU}{ReLU}

\usepackage[
    maxbibnames=2,
    backend=biber,
    bibstyle=ieee,
    citestyle=numeric-comp,
    doi=false,
    url=false,
    isbn=false,
    sorting=none]{biblatex}
\DeclareFieldFormat[misc]{title}{%
    \iffieldequalstr{entrykey}{valentini2017noisy}
        {\mkbibquote{#1\isdot}}
        {\mkbibemph{#1}}}
\AtBeginBibliography{\setlength{\emergencystretch}{3em}}

\newcolumntype{P}[1]{>{\raggedleft\arraybackslash}p{#1}}

\usepackage[hidelinks]{hyperref}
\usepackage[
    acronym,      %
    toc,          %
    nonumberlist  %
]{glossaries}
\newcommand{\DeclareAcronym}[2]{\newacronym{#1}{#1}{#2}}
\newcommand{\ac}[1]{\gls*{#1}}

\newcommand{\acpl}[1]{\glspl*{#1}}
\newcommand{\Acpl}[1]{\Glspl*{#1}}

\newcommand{\dsru}{DuSpaR}
\newacronym{sru}{SpaR}{Sparsifying Recurrent Unit}
\DeclareAcronym{KWS}{keyword spotting}
\DeclareAcronym{SLU}{spoken language understanding}
\DeclareAcronym{ASR}{automatic speech recognition}
\DeclareAcronym{SE}{speech enhancement}
\DeclareAcronym{RNN}{recurrent neural network}
\DeclareAcronym{LSTM}{Long Short-Term Memory}
\DeclareAcronym{GRU}{Gated Recurrent Unit}
\DeclareAcronym{MVM}{matrix-vector multiplication}
\DeclareAcronym{MAC}{multiply-accumulate}
\DeclareAcronym{\dsru}{Dual-state Sparsifying Recurrent Unit}
\DeclareAcronym{ReLU}{Rectified Linear Unit}

\usepackage{page3figure}
\PlaceTaskFigureOnPageThree{%
    \centering
    \includegraphics[width=\textwidth]{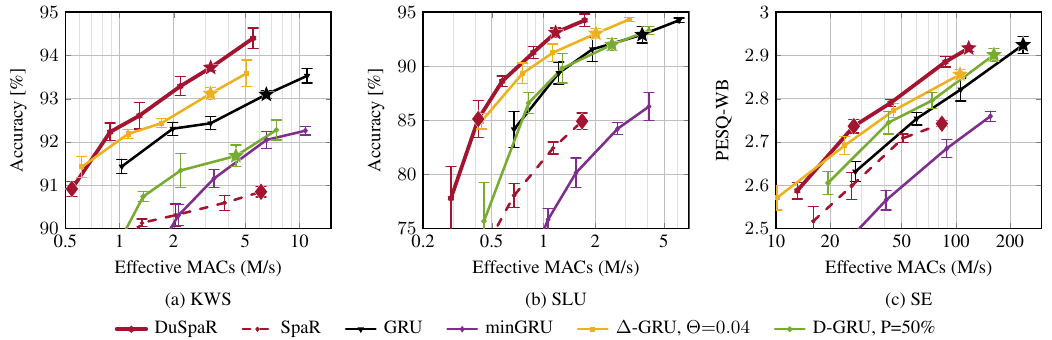}
    \vspace{-2\baselineskip}
    \caption{Task performance versus total effective computation for
        (a) KWS, (b) SLU, and (c) SE.
        Error bars indicate standard deviation over five runs.
        Each curve shows results for one recurrent model at the layer widths described in Section~\ref{sec:ablation}.
        Stars (\(\bigstar\)) mark the DuSpaR, GRU, $\Delta$-GRU, and D-GRU configurations in Table~\ref{tab:kws_slu_se},
        which have similar parameter counts within each task.
        Diamonds (\(\blacklozenge\)) mark the DuSpaR--SpaR pairs used to evaluate the effectiveness of dual-state recurrence 
        by comparing computational cost at similar task performance.}
    \label{fig:task_tradeoffs}
    \vspace{-1\baselineskip}
}

\title{
    DuSpaR: Dual-State Sparsifying Recurrent Unit with Feedback Modulation for Compute-Efficient Speech Processing
    }

\name{Zixiao Li, Sheng Zhou, Longbiao Cheng, Shih-Chii Liu\thanks{This work was partially funded by the Swiss National Science Foundation project CA-DNNEdge (208227).}}
\address{Institute of Neuroinformatics, University of Zurich and ETH Zurich, Zurich, Switzerland}

\begin{document}
\ninept
\maketitle
\ActivatePageThreeFigure

\begin{abstract}
    We introduce the Dual-state Sparsifying Recurrent Unit~(\dsru) as a computationally efficient building block for speech processing models on resource-constrained edge devices.
    It employs dual-state recurrence to modulate its input vectors in a stateful feedback loop.
    Its recurrent cells sparsify the input vector operand involved in matrix-vector multiplication using ReLU activation.
    By skipping the zero entries dynamically, inference-time savings in multiply-accumulate operations and weight memory fetches can be achieved.
    We evaluate \dsru \ on three speech tasks: keyword spotting~(KWS) on the Google Speech Commands dataset,
    spoken language understanding~(SLU) on the Fluent Speech Commands dataset,
    and speech enhancement~(SE) on the Voice Bank + Demand~(VBD) dataset.
    At similar parameter counts, \dsru \ requires 51.0\% and 68.4\% less computation than Gated Recurrent Unit (GRU) on KWS and SLU, respectively, while achieving higher accuracy,
    and 50.1\% less computation on SE while maintaining similar quality.
    At comparable computational cost and across a range of model sizes,
    \dsru \ also achieves higher KWS/SLU accuracy and better SE quality
    than other sparsity-aware recurrent models.
    Ablation studies show that compared with the single-state recurrence baseline, dual-state recurrence reduces the effective compute by 3.1$\times$ to 11.3$\times$ at similar task performance.
\end{abstract}


\section{Introduction}
\label{sec:intro}

Neural networks for speech processing are widely employed in mobile platforms including smartphones, smart glasses, wearables, and hearables.
They support voice interaction and improve speech intelligibility,
performing tasks such as \ac{KWS}, \ac{ASR}, \ac{SLU}, and \ac{SE}.
Compared with remote processing in the cloud,
deploying these networks locally on the device offers lower latency, higher reliability, and better user privacy~\cite{2019-ICASSP-He-StreamingASR}.
However, the limited compute and memory resources available on device create a significant challenge for their deployment~\cite{2020-Interspeech-Fedorov-TinyLSTM},
necessitating the design of compact and efficient networks.

\Acpl{RNN} such as \ac{GRU}~\cite{2014-NIPS-Chung-GRU} and \ac{LSTM}~\cite{1997-NECO-Hochreiter-LSTM} are well suited to streaming speech processing because they encode the input context as a fixed-size recurrent state,
whose storage does not grow with the duration of the input stream~\cite{2024-ICASSP-Rong-GTCRN, 2024-Interspeech-Cheng-DGRU}.
New recurrent models such as LRU~\cite{2023-ICML-Orvieto-LRU}, xLSTM~\cite{2024-NIPS-Beck-xLSTM}, S4~\cite{2022-ICLR-Gu-S4}, and Mamba~\cite{2024-ICML-Dao-Mamba2}
show continual improvement in performance and training efficiency for sequence modeling.
Nevertheless, these recurrent layers perform dense \ac{MVM} at every inference step,
leading to a large number of \ac{MAC} operations and weight memory fetches.
Therefore, reducing the inference cost of recurrent layers without compromising task performance is an important target for on-device speech-processing networks.

\begin{figure}[t]
    \centering
    \includegraphics[width=\linewidth]{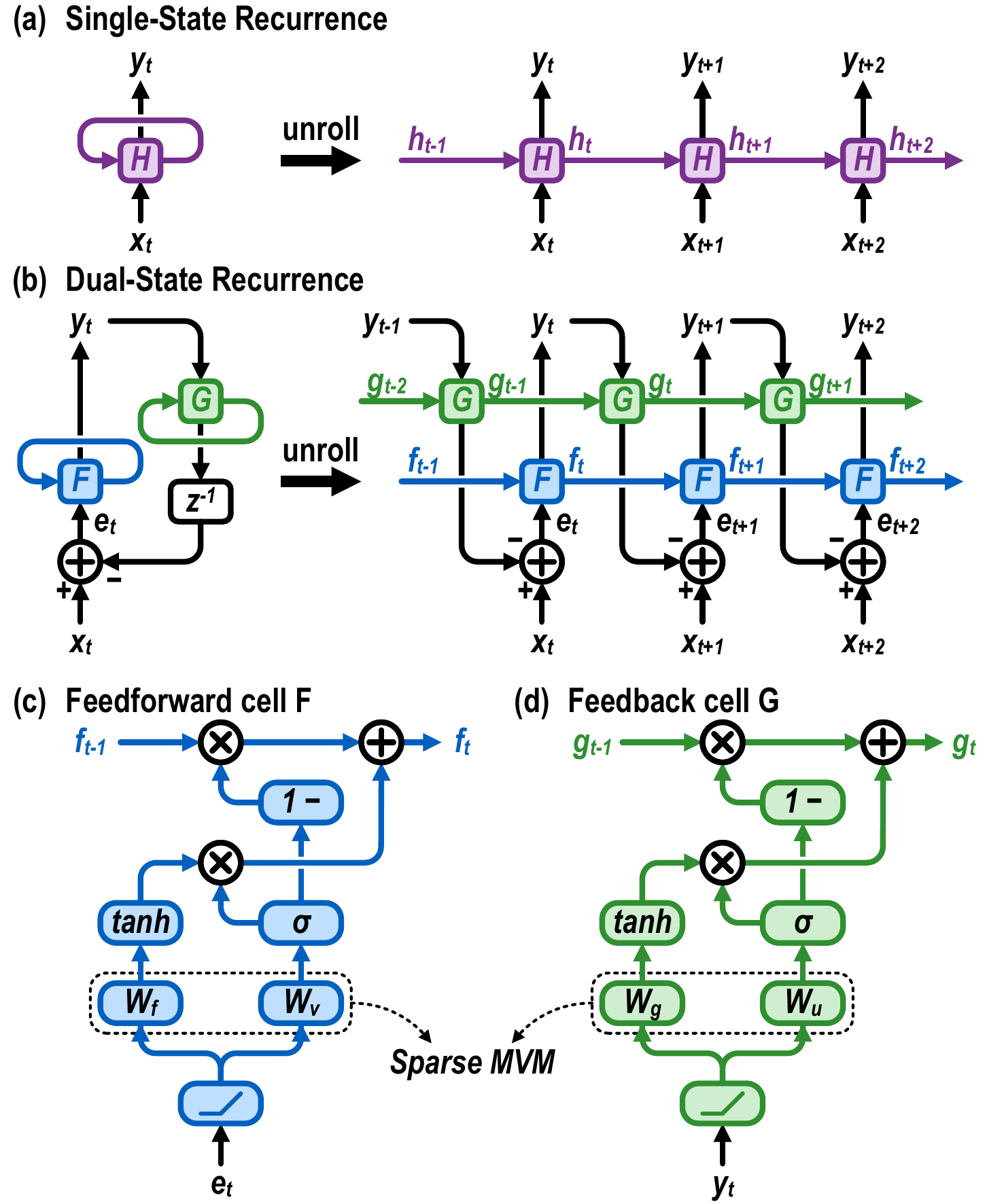}
    \vspace{-2\baselineskip}
    \caption{
        Illustration of the proposed Dual-state Sparsifying Recurrent Unit~(\dsru).
        (a)~Conventional single-state recurrence.
        (b)~Proposed dual-state recurrence adopted in \dsru.
        (c)~Implementation of the feedforward recurrent cell $\bm{F}$.
        (d)~Implementation of the feedback recurrent cell $\bm{G}$.
        All vector operands in matrix-vector multiplication~(MVM) are sparsified by ReLU.
    }
    \label{fig:arch}
    \vspace{-1\baselineskip}
\end{figure}

Several methods can be employed to optimize the inference cost of \acpl{RNN}.
Quantization~\cite{2021-Interspeech-Fasoli-QuantizedLSTM} of weights and activations reduces the unit cost per \ac{MAC}.
Pruning~\cite{2019-FPGA-Cao-BBS} provides static sparsification of weights,
allowing zero-weight \acpl{MAC} to be skipped.
Dynamic computation~\cite{2017-ICML-Neil-DeltaNetwork, 2023-ICLR-Subramoney-EGRU, 2024-Interspeech-Cheng-DGRU, 2018-ICLR-Campos-SkipRNN, 2022-TPAMI-Han-DynamicNetworkSurvey} instead relies on dynamic sparsification of activations,
which enables input-dependent skipping of \acpl{MAC} and memory fetches during inference.
These models have demonstrated significant throughput improvement and energy reduction when deployed on custom hardware~\cite{2020-JETCAS-Gao-EdgeDRNN, 2025-JSSC-Zhou-SLU}.

In this work, we introduce \ac{\dsru}, a novel recurrent building block~(Fig.~\ref{fig:arch}) improving the trade-off between task performance and compute cost compared to previous dynamic computation methods.
Our contributions are as follows:
\begin{enumerate}
    \item
          A dual-state recurrence scheme~(Section~\ref{sec:method-dual-state}) that modulates the input temporal sequence through a stateful feedback loop.
          Ablation studies~(Section~\ref{sec:results}) show that the feedback reduces the number of \acpl{MAC} at similar task performance compared to conventional single-state recurrence.
    \item
          A sparsifying recurrent cell~(Section~\ref{sec:method-sparse-cell}) that uses ReLU to sparsify the vector operand in \ac{MVM} while maintaining a bounded state using hyperbolic tangent activation.
          Coupling two such cells through dual-state recurrence results in the proposed \ac{\dsru}.
    \item
          Comprehensive evaluation of \ac{\dsru} on multiple speech tasks~(Section~\ref{sec:setup}), covering both classification and regression.
          Comparisons across model sizes show that \ac{\dsru} outperforms both dense and sparse recurrent models, achieving better task performance at comparable computational cost.
\end{enumerate}

\section{Methods}
\label{sec:method}

In the following subsections, activation and state vectors are denoted by lowercase bold letters such as $\bm{x}$ and $\bm{y}$,
weight matrices by uppercase bold letter $\bm{W}$,
and bias vectors by lowercase bold letter $\bm{b}$.
Nonlinear functions such as $\tanh$ and $\ReLU$ are applied elementwise, and $\odot$ denotes elementwise product of two vectors.

\subsection{Dual-State Recurrence}
\label{sec:method-dual-state}

Fig.~\ref{fig:arch}\,(a) illustrates a conventional single-state recurrent layer unrolled over time, with inputs $[\bm{x}_t]_{t=1}^T$, outputs $[\bm{y}_t]_{t=1}^T$, and recurrent states $[\bm{h}_t]_{t=1}^T$.
At each time step $t$, the recurrent cell $\bm{H}$ computes the new state $\bm{h}_t$ from the current input $\bm{x}_t$ and the previous state $\bm{h}_{t-1}$ as
\(
\bm{h}_t = \bm{H}(\bm{x}_t, \bm{h}_{t-1})
\).
In a multi-layer network, multiple recurrent cells are stacked in a feedforward manner.
Therefore, information flows \textit{one-way} from the inputs $[\bm{x}_t]_{t=1}^T$ to the outputs $[\bm{y}_t]_{t=1}^T$.

Similar to models~\cite{han2018deep} inspired by the prevalent local recurrent processing in brains~\cite{1991-Cortex-Felleman-Reciprocal},
we introduce the dual-state recurrence shown in Fig.~\ref{fig:arch}\,(b).
The two recurrent cells $\bm{F}$ and $\bm{G}$ are coupled in a feedback loop,
with separate recurrent states $[\bm{f}_t]_{t=1}^T$ and $[\bm{g}_t]_{t=1}^T$.
$\bm{F}$ sends \textit{feedforward} information from $[\bm{e}_t]_{t=1}^T$ to $[\bm{y}_t]_{t=1}^T$,
while $\bm{G}$ sends \textit{feedback} information from $[\bm{y}_t]_{t=1}^T$ to $[\bm{e}_t]_{t=1}^T$.
$\bm{e}_t$ is computed by subtracting the previous feedback state $\bm{g}_{t-1}$ from the input $\bm{x}_t$.
In other words, the input sequence $[\bm{x}_t]_{t=1}^T$ is \textit{modulated} by the feedback states $[\bm{g}_t]_{t=1}^T$.
We hypothesize that this stateful feedback modulation of the inputs enhances the expressiveness of the model,
and empirically confirm its effectiveness through an ablation study~(Section~\ref{sec:results}).
Mathematically, the dual-state recurrence is described by Eqs.~\ref{eq:dual-state-error}--\ref{eq:dual-state-feedback},
where $\bm{b}_g$ and $\bm{b}_f$ are trainable bias vectors.
\begin{align}
    \label{eq:dual-state-error}
    \bm{e}_t
     & = \bm{x}_t - (\bm{g}_{t-1} + \bm{b}_g)
     & \;\rhd
     & \text{ } \mathcbox[natblue!20]{\text{Feedforward cell}} \text{ input}        \\
    \label{eq:dual-state-feedforward}
    \bm{f}_t
     & = \bm{F}(\bm{e}_t, \bm{f}_{t-1})
     & \;\rhd
     & \text{ } \mathcbox[natblue!20]{\text{Feedforward cell}} \text{ state update} \\
    \label{eq:dual-state-output}
    \bm{y}_t
     & = \bm{f}_t + \bm{b}_f
     & \;\rhd
     & \text{ } \mathcbox[natgreen!20]{\text{Feedback cell}} \text{ input}          \\
    \label{eq:dual-state-feedback}
    \bm{g}_t
     & = \bm{G}(\bm{y}_t, \bm{g}_{t-1})
     & \;\rhd
     & \text{ } \mathcbox[natgreen!20]{\text{Feedback cell}} \text{ state update}
\end{align}

\subsection{Sparsifying Recurrent Cell}
\label{sec:method-sparse-cell}

To dynamically reduce the \ac{MAC} operations and weight memory fetches,
we employ the sparsifying recurrent cell shown in Fig.~\ref{fig:arch}\,(c) for $\bm{F}$ and Fig.~\ref{fig:arch}\,(d) for $\bm{G}$.
The cell is adapted from minGRU~\cite{2018-ICLR-Martin-GILR, 2024-ArXiv-Feng-minGRU},
which involves two \acpl{MVM}, one for computing the update gate, and the other for the update candidate.
We sparsify both \acpl{MVM} by using ReLU to zero out the negative elements of the input vector.
In addition, we apply $\tanh$ activation to the update candidate to ensure that the dense state vector stays bounded even for long sequences.
The sparsifying recurrent cell for $\bm{F}$ is given by Eqs.~\ref{eq:feedforward-input}--\ref{eq:feedforward-update},
where $\sigma$ is the sigmoid activation and the differences with minGRU are highlighted in purple.
\begin{align}
    \label{eq:feedforward-input}
    \bm{e}^+_{t}
     & = \mathcbox[natpurple!20]{\ReLU}(\bm{e}_t)
     & \;\rhd
     & \text{ Input sparsification}                                                      \\
    \label{eq:feedforward-gate}
    \bm{v}_t
     & = \sigma(\bm{W}_v \bm{e}^+_t + \bm{b}_v)
     & \;\rhd
     & \text{ Update gate}                                                               \\
    \label{eq:feedforward-candidate}
    \tilde{\bm{f}}_t
     & = \mathcbox[natpurple!20]{\tanh}(\bm{W}_f \bm{e}^+_t)
     & \;\rhd
     & \text{ Update candidate}                                                          \\
    \label{eq:feedforward-update}
    \bm{f}_t
     & = (1-\bm{v}_t)\odot \bm{f}_{t-1} + \bm{v}_t\odot \tilde{\bm{f}}_t
     & \;\rhd
     & \text{ Gated state update}
\end{align}
Similarly, the sparsifying recurrent cell for $\bm{G}$ is given by Eqs.~\ref{eq:feedback-input}--\ref{eq:feedback-update}.
\begin{align}
    \label{eq:feedback-input}
    \bm{y}^+_{t}
     & = \mathcbox[natpurple!20]{\ReLU}(\bm{y}_t)
     & \;\rhd
     & \text{ Input sparsification}                                                      \\
    \label{eq:feedback-gate}
    \bm{u}_t
     & = \sigma(\bm{W}_u \bm{y}^+_t + \bm{b}_u)
     & \;\rhd
     & \text{ Update gate}                                                               \\
    \label{eq:feedback-candidate}
    \tilde{\bm{g}}_t
     & = \mathcbox[natpurple!20]{\tanh}(\bm{W}_g \bm{y}^+_t)
     & \;\rhd
     & \text{ Update candidate}                                                          \\
    \label{eq:feedback-update}
    \bm{g}_t
     & = (1-\bm{u}_t)\odot \bm{g}_{t-1} + \bm{u}_t\odot \tilde{\bm{g}}_t
     & \;\rhd
     & \text{ Gated state update}
\end{align}
Our sparsifying recurrent cell is based on minGRU because its recurrent states are only involved in elementwise operations instead of \ac{MVM} as in \ac{GRU} or \ac{LSTM}.
This allows us to sparsify the \acpl{MVM}, which dominate the overall computational cost,
without obscuring the information flow through time by keeping the recurrent states as dense vectors.

For a \ac{\dsru} layer with input dimension $N$ and output dimension $M$,
there are $4NM+2(N+M)$ trainable parameters in total, including
\(
\bm{W}_f, \bm{W}_g, \bm{W}_u, \bm{W}_v,
\bm{b}_f, \bm{b}_g, \bm{b}_u, \bm{b}_v
\).
With elementwise operations for the gated state updates~(Eqs.~\ref{eq:feedforward-update} and \ref{eq:feedback-update}) included,
the number of effective \acpl{MAC} per time step is
reduced from $4NM+2(N+M)$ to
$
2(o_e+o_y)NM + 2(N+M)
$,
where $o_e$ and $o_y$ are the average occupancies of $\bm{e}^+_t$~(Eq.~\ref{eq:feedforward-input}) and $\bm{y}^+_t$~(Eq.~\ref{eq:feedback-input}), respectively.
The \ac{MAC} reduction ratio is approximately $(o_e+o_y)/2$.
Likewise, the weight memory fetches per time step are reduced from $4NM$ to $2(o_e+o_y)NM$,
by a ratio of $(o_e+o_y)/2$.

\subsection{Related Work}
\label{sec:method-related-works}

\paragraph{Efficient RNN}
The fine-grained activation sparsity in \ac{\dsru} is complementary to other techniques for reducing \ac{RNN} inference cost,
including quantization~\cite{2021-Interspeech-Fasoli-QuantizedLSTM},
pruning~\cite{2019-FPGA-Cao-BBS}, and coarse-grained frame skipping~\cite{2018-ICLR-Campos-SkipRNN}.
Combining these techniques further reduces the inference cost of speech processing \acpl{RNN}~\cite{2022-TNNLS-Gao-Spartus}.

\paragraph{Activation Sparsity}
Several prior works~\cite{2017-ICML-Neil-DeltaNetwork, 2024-Interspeech-Cheng-DGRU,2023-ICLR-Subramoney-EGRU} also exploit fine-grained activation sparsity for efficient recurrent inference.
Delta Network~\cite{2017-ICML-Neil-DeltaNetwork} skips entries of the input and state vectors whose activation change across two time steps is below a set threshold.
EGRU~\cite{2023-ICLR-Subramoney-EGRU} introduces a stateful gating mechanism with a trainable threshold for sparse, event-based updates.
Dynamic Gated \ac{RNN}~\cite{2024-Interspeech-Cheng-DGRU} skips the update of a fixed fraction of recurrent neurons based on their update-gate magnitudes.
These methods introduce sparsity by modifying or approximating an underlying dense model such as \ac{GRU}.
In contrast, our proposed \ac{\dsru} is intrinsically sparse and achieves better trade-offs between task performance and effective \ac{MAC} operations at comparable model sizes~(Section~\ref{sec:results}).

\paragraph{ReLU RNN}
The ReLU activation in \ac{\dsru} has been used in other recurrent models.
IRNN~\cite{2015-arXiv-Le-IRNN} produces sparse recurrent states using the ReLU nonlinearity,
while \ac{\dsru} retains dense recurrent states and applies the sparsification only to the recurrent cell inputs.
Li-\ac{GRU}~\cite{2018-TETCI-Ravanelli-LiGRU} simplifies the \ac{GRU} architecture by removing the reset gate and replacing the $\tanh$ activation with ReLU in calculating the update candidate.
However, the update candidate is only involved in elementwise operations,
so the computational savings are negligible.

\section{Experimental Setup}
\label{sec:setup}
In  Section~\ref{sec:tasks}, we evaluate \ac{\dsru} on three speech tasks with different objectives and temporal context windows:
KWS, SLU, and SE.
For each task, we use a specific GRU-based architecture and conduct comparison experiments
by replacing its GRU modules with \ac{\dsru} and other recurrent models described in Section~\ref{sec:ablation}.

\subsection{Tasks}
\label{sec:tasks}

\paragraph{Keyword Spotting}
KWS detects target keywords using word-level context spanning hundreds of milliseconds.
We use the Google Speech Commands Dataset (GSCD)~\cite{2018-ArXiv-Warder-GSCD},
which contains 105,000 one-second utterances covering 35 keyword classes.
All recordings from GSCD are sampled at 16\,kHz and processed using 32-ms Hann windows with a 16-ms hop.
Each frame's spectrogram is then computed using a 512-point Fast Fourier Transform (FFT) and mapped to 64 mel bands spanning from 50\,Hz to 8\,kHz.
The mel-band features are log-transformed and standardized using training-set statistics. The KWS model uses two GRU layers and one fully connected (FC) classification layer.
We train the model for up to 100 epochs with batch size 64 and standard cross-entropy (CE) loss.

\paragraph{Spoken Language Understanding}
SLU infers user intent at the sentence level from a context window spanning several seconds.
We use the Fluent Speech Commands Dataset (FSCD)~\cite{2019-Interspeech-Lugosch-FSCD},
which contains 19 hours of English smart-home commands recordings from 97 speakers,
labeled with 31 intents defined by action-object-location combinations.
We use the identical KWS audio front-end for SLU, and a model with two GRU layers and one FC layer~\cite{2025-JSSC-Zhou-SLU}.
To reduce computation, two temporal pooling layers, each with a window size of 4, are applied to the GRU outputs.
We train the model for 200 epochs with batch size 64 and the Connectionist
Temporal Classification (CTC) loss~\cite{2006-ICML-Schmidhuber-CTC}.
During the first 100 epochs, we supplement the CTC loss with an auxiliary
CE loss to improve convergence.

\paragraph{Speech Enhancement}
SE estimates clean speech from noisy inputs lasting up to tens of seconds. We use the Voice Bank + DEMAND (VBD) dataset~\cite{valentini2017noisy}.
The training set mixes clean speech from 28 of the 30 speakers in the Voice Bank dataset~\cite{veaux2013voice} with 10 of the 15 noise types
in the DEMAND dataset~\cite{thiemann2013diverse} at four signal-to-noise ratio (SNR) levels: 0, 5, 10, and 15\,dB.
For testing, clean speech from the remaining two speakers is mixed with the remaining five noise types at SNRs of 2.5, 7.5, 12.5, and 17.5\,dB.
To target 2\,ms ultra-low enhancement latency~\cite{2025-ICASSP-Cheng-SlowFast}, we sample audio at 16\,kHz and use 32-sample (2\,ms) frames with a 16-sample (1\,ms) hop.

The SE model directly processes these time-domain frames using an encoder-mask-decoder architecture~\cite{2025-ICASSP-Cheng-SlowFast}.
A linear encoder maps each frame to learned features.
Four GRU layers followed by one FC layer estimate a mask that multiplies
the encoder features element-wise.
A linear decoder converts the masked features to waveform frames for audio reconstruction
by Hann-windowed Overlap-and-Add (OLA).
We train for 280 epochs with batch size 32,
and only the mean-squared error (MSE) loss is used for the first 200 epochs.
For the last 80 epochs, we adopt the loss function and weighting factors from~\cite{2025-ICASSP-Cheng-SlowFast}. This loss function combines MSE,
perceptual-based loss terms from~\cite{2018-LSP-Martin-PerceptualLoss} and~\cite{2018-ICASSP-Zhang-PerceptualLoss}, and a negative Scale-Invariant Signal-to-Noise Ratio (SI-SNR) loss term.
We evaluate SE performance using PESQ-WB~\cite{2001-ICASSP-Rix-PESQ}, STOI~\cite{2011-TASLP-Taal-STOI}, and scale-invariant signal-to-distortion ratio (SI-SDR)~\cite{2019-ICASSP-Roux-SISDR}.

\begin{table*}[t]
\centering
\fontsize{9}{10.45}\selectfont
\caption{KWS and SLU accuracy, 
SE metrics, and computational cost of recurrent models with comparable parameter counts within each task. 
Entries with $\pm$ report the mean and standard deviation over five runs.
Dashes indicate entries that are not applicable.}
\label{tab:kws_slu_se}
\setlength{\tabcolsep}{3pt}
\renewcommand{\arraystretch}{1}
\newcommand{\combinedtabnote}[1]{\smash{\textsuperscript{#1}}}
\makeatletter
\newcommand{\combinedrowhighlight}{%
  \noalign{\begingroup
    \color{natgreen!15}%
    \dimen@=\dimexpr\ht\@arstrutbox+\dp\@arstrutbox\relax
    \hrule height\dimen@
    \kern-\dimen@
  \endgroup}%
}
\makeatother
\begin{tabular*}{\textwidth}{@{\extracolsep{\fill}}*{11}{c}@{}}
\toprule
Task & Model & \makecell{Para.\\(K)} & \makecell{Layer\\Width} & \makecell{Dense\\MACs (M/s)\combinedtabnote{1}} & \makecell{Effective\\MACs (M/s)\combinedtabnote{2}\,$\downarrow$} & \makecell{Occ.\combinedtabnote{3}\,$\downarrow$\\(\%)} & \makecell{Accuracy\,$\uparrow$\\(\%)} & PESQ-WB\,$\uparrow$ & \makecell{STOI\,$\uparrow$\\(\%)} & \makecell{SISDR\,$\uparrow$\\(dB)} \\
\midrule
\multirow{4}{*}{\textit{KWS}} & GRU & 105.9 & 96 & 6.55 & 6.55 & 100 & 93.10 $\pm$ 0.10 & - & - & - \\
 & $\Delta$-GRU\combinedtabnote{4} & 105.9 & 96 & 6.55 & \textbf{3.21} & \textbf{49.0} & 93.12 $\pm$ 0.14 & - & - & - \\
 & D-GRU\combinedtabnote{5} & 105.9 & 96 & 6.55 & 4.43 & 67.7 & 91.68 $\pm$ 0.25 & - & - & - \\
\combinedrowhighlight
 & \ac{\dsru} & 103.7 & 128 & 6.42 & \textbf{3.21} & 50.0 & \textbf{93.72} $\pm$ 0.06 & - & - & - \\
\midrule
\multirow{4}{*}{\textit{SLU}} & GRU & 105.6 & 96 & 3.74 & 3.74 & 100 & 92.92 $\pm$ 0.77 & - & - & - \\
 & $\Delta$-GRU\combinedtabnote{4} & 105.6 & 96 & 3.74 & 2.02 & 53.8 & 93.03 $\pm$ 0.48 & - & - & - \\
 & D-GRU\combinedtabnote{5} & 105.6 & 96 & 3.74 & 2.50 & 66.8 & 92.00 $\pm$ 0.54 & - & - & - \\
\combinedrowhighlight
 & \ac{\dsru} & 103.3 & 128 & 3.08 & \textbf{1.18} & \textbf{38.3} & \textbf{93.12} $\pm$ 0.43 & - & - & - \\
\midrule
\multirow{5}{*}{\textit{SE}} & noisy & - & - & - & - & - & - & 1.97 & 92.10 & 8.45 \\
 & GRU & 239.5 & 96 & 236.78 & 236.78 & 100 & - & \textbf{2.92} $\pm$ 0.02 & \textbf{94.02} $\pm$ 0.05 & \textbf{18.43} $\pm$ 0.05 \\
 & $\Delta$-GRU\combinedtabnote{4} & 239.5 & 96 & 236.78 & \textbf{105.58} & \textbf{44.6} & - & 2.86 $\pm$ 0.01 & 93.76 $\pm$ 0.02 & 18.34 $\pm$ 0.05 \\
 & D-GRU\combinedtabnote{5} & 239.5 & 96 & 236.78 & 162.98 & 68.8 & - & 2.90 $\pm$ 0.01 & 93.91 $\pm$ 0.09 & 18.31 $\pm$ 0.06 \\
\combinedrowhighlight
 & \ac{\dsru} & 238.8 & 116 & 236.41 & 118.04 & 49.9 & - & \textbf{2.92} $\pm$ 0.01 & 93.82 $\pm$ 0.13 & 18.34 $\pm$ 0.03 \\
\bottomrule
\end{tabular*}

\par\smallskip
\begin{minipage}{\textwidth}
\fontsize{9}{10.45}\selectfont
\raggedright
\setlength{\parskip}{0pt}
\hangindent=0.6em\hangafter=1
\noindent\makebox[0.6em][l]{\combinedtabnote{1}}All-layer MAC cost without sparsity-based skipping; M/s denotes millions of MACs per second.
\par
\hangindent=0.6em\hangafter=1
\noindent\makebox[0.6em][l]{\combinedtabnote{2}}All-layer MAC cost after accounting for sparsity-based skipping.
\par
\hangindent=0.6em\hangafter=1
\noindent\makebox[0.6em][l]{\combinedtabnote{3}}Occ. is shorthand for occupancy, where occupancy $= (\text{Effective MACs}/\text{Dense MACs}) \times 100\%$.
\par
\hangindent=0.6em\hangafter=1
\noindent\makebox[0.6em][l]{\combinedtabnote{4}}$\Delta$-GRU uses $\Theta=0.04$ for all tasks.
\quad\makebox[0.6em][l]{\combinedtabnote{5}}D-GRU uses $P=50\%$ for all tasks.
\end{minipage}
\par\vspace{-8pt}
\end{table*}

\subsection{Recurrent Model Comparison and Ablation}
\label{sec:ablation}
We compare \ac{\dsru} with both dense and sparsity-aware recurrent models to evaluate their trade-offs between task performance and computational cost.
For each network described in Section~\ref{sec:tasks},
we replace its \ac{GRU} layers with \ac{\dsru} or other recurrent models listed below while keeping the remaining network structure fixed.
For the dense variants, apart from the default \ac{GRU} layer,
we include minGRU~\cite{2018-ICLR-Martin-GILR,2024-ArXiv-Feng-minGRU} due to its connection with \ac{\dsru}~(Section~\ref{sec:method-sparse-cell}).
For the sparse variants, we include $\Delta$-GRU as a representative of Delta Networks~\cite{2017-ICML-Neil-DeltaNetwork}
and D-GRU as a representative of Dynamic Gated \ac{RNN}~\cite{2024-Interspeech-Cheng-DGRU}.
In addition, as an ablation study on the contribution of the dual-state recurrence~(Section~\ref{sec:method-dual-state}),
we also consider a variant of \ac{\dsru} without the feedback path.
We call this variant \ac{sru} and it is equivalent to retaining only the feedforward path of \ac{\dsru} following
Eq.~\ref{eq:feedforward-input}-\ref{eq:feedforward-update} with $\bm{e}_t = \bm{x}_t$.
We also experimented with IRNN~\cite{2015-arXiv-Le-IRNN} and Li-GRU~\cite{2018-TETCI-Ravanelli-LiGRU},
but exploding gradients prevented successful training in our SLU and SE experiments.
We therefore excluded both models from our final comparison.

We note that $\Delta$-GRU and D-GRU use the delta threshold $\Theta$ and update ratio $P$, respectively, to control the trade-off between task performance and computational cost.
Suitable values for these hyperparameters depend on the task, model, and desired trade-off. We sweep $\Theta \in \{0.02, 0.04, 0.06\}$ for $\Delta$-GRU and $P \in \{25\%, 50\%\}$ for D-GRU,
and select $\Theta = 0.04$ and $P = 50\%$ because they provide favorable trade-offs across all tasks. Although other values may yield better trade-offs for specific task and model configurations,
identifying them can require substantial tuning effort, representing a potential limitation of both algorithms.

We define the layer width as the output dimension of each recurrent layer.
For \ac{\dsru}, it is the dimension of the feedforward state $\bm{f}_t$ in Eq.~\ref{eq:dual-state-feedforward}.
For KWS and SLU, we use a common layer-width grid of \{32, 48, 64, 96, 128\}, extended to 192 for DuSpaR and to \{192, 256\} for 
SpaR and minGRU. For SE, the common grid is \{32, 48, 64, 96\}, with additional widths of 116 for DuSpaR and 128 for SpaR and minGRU. 
Because parameter counts differ across recurrent architectures at the same width, 
we include these larger widths so that all models cover similar ranges of parameter counts within each task.
The effective MAC cost across all layers is reported in millions of MACs per second (M/s).
This count excludes operations skipped through input sparsity in \ac{\dsru}, \ac{sru}, and $\Delta$-GRU, or output sparsity in D-GRU.

\section{Results}
\label{sec:results}

Fig.~\ref{fig:task_tradeoffs} plots task performance against effective MAC cost for the three tasks.
Curves toward the upper left indicate better performance with fewer computes.
The \ac{sru} achieves a better performance--computation trade-off than minGRU on SLU and SE, but still worse than GRU.
Coupling two \ac{sru} cells through feedback modulation to form \ac{\dsru} improves computational efficiency.
For the \ac{\dsru}--\ac{sru} pairs marked by diamonds (\(\blacklozenge\)) in Fig.~\ref{fig:task_tradeoffs},
\ac{\dsru} reduces effective MAC cost by
approximately 11.3$\times$, 4.0$\times$, and 3.1$\times$
on KWS, SLU, and SE, respectively, at comparable task performance.
Across all three tasks, \ac{\dsru} achieves a substantially better performance-computation trade-off than GRU and D-GRU,
as reflected by its curves lying further toward the upper left in Fig.~\ref{fig:task_tradeoffs}.
While $\Delta$-GRU remains competitive for small models, \ac{\dsru} holds a clear advantage for larger models.

Table~\ref{tab:kws_slu_se} compares GRU, $\Delta$-GRU, D-GRU, and \ac{\dsru} at comparable parameter counts within each task.
These configurations are marked by stars (\(\bigstar\)) in Fig.~\ref{fig:task_tradeoffs}.
We use a larger layer width for \ac{\dsru} so that its parameter count is close to, but slightly below, that of the other three models.
\ac{\dsru} combines the highest mean
accuracy on KWS and SLU with the lowest effective MAC costs.
On SLU, it achieves 93.12\% accuracy, 
with an effective MAC cost that is lower by 68.4\%, 41.6\%, and 52.8\% relative to GRU, $\Delta$-GRU,
and D-GRU, respectively.
Its 38.3\% occupancy is the lowest among the four models.
On KWS, \ac{\dsru} and $\Delta$-GRU share the minimum cost of
3.21~MMACs/s with occupancies near 50\%, but \ac{\dsru}'s 93.72\%
accuracy is 0.6 percentage points higher. For SE, \ac{\dsru} achieves the highest PESQ-WB score, with a small decrease in STOI and SI-SDR relative to GRU, but GRU requires approximately $2\times$ more MACs/s.
Although $\Delta$-GRU has a lower effective MAC cost for SE, \ac{\dsru} provides higher PESQ-WB and STOI scores.

\section{Conclusion}

We introduce \ac{\dsru}, a recurrent unit that combines coupled feedforward and feedback states with ReLU-based vector sparsification 
to reduce computes for on-device inference of speech processing networks. 
Results show that \ac{\dsru} achieves 
better performance--computation trade-offs on KWS, SLU, and SE tasks than the previous state-of-the-art models without the feedback recurrent cell.
Future work includes experiments to further understand the role of the dual-state recurrence in improving the trade-off, extending the evaluation to larger models and tasks with longer temporal dependencies, e.g., long-form speech recognition and language modeling; and exploring applications in other modalities.

\printbibliography[heading=bibnumbered,title={REFERENCES}]

\end{document}